\documentclass[a4paper,11pt]{article}
\usepackage{tikz-feynman}
\usepackage{tikz}
\tikzfeynmanset{compat=1.0.0}
\usetikzlibrary{shapes,arrows,positioning,automata,backgrounds,calc,er,patterns}
\usepackage{jcappub} 

\usepackage[T1]{fontenc} 
\usepackage{array,makecell}
\usepackage{mathtools}
\usepackage{graphicx}
\usepackage{enumerate}
\usepackage{subfig}
\usepackage{amsmath}
\usepackage{cite}

\usepackage[utf8]{inputenc}

\usepackage[T1]{fontenc}
\usepackage{multirow}

\usepackage[mathscr]{eucal}
\usepackage[Symbolsmallscale]{upgreek}
\usepackage{xcolor}
\usepackage{array}

\title{\boldmath 
Probing Axions via Light Circular Polarization and Event Horizon Telescope 
}

\author[a,b]{Soroush Shakeri,}
\author[c,d,e]{Fazlollah Hajkarim}

\affiliation[a]{Department of Physics, Isfahan University of Technology, Isfahan 84156-83111, Iran}
\affiliation[b]{ICRANet, Piazza della Repubblica 10, I-65122 Pescara, Italy}
\affiliation[c]{ Dipartimento di Fisica e Astronomia, Università degli Studi di Padova, \\ Via Marzolo 8, 35131 Padova, Italy
}

\affiliation[d]{Istituto Nazionale di Fisica Nucleare (INFN), Sezione di Padova, \\ Via Marzolo 8, 35131 Padova, Italy}
\affiliation[e]{Department of Physics and Astronomy, University of Oklahoma, Norman, OK 73019, U.S.A.}

\emailAdd{s.shakeri@iut.ac.ir}
\emailAdd{hajkarim@pd.infn.it}

\abstract{The impact of axion-like particles on the light polarization around the horizon of supermassive black hole (SMBH) is discussed in the light of the latest polarization measurement of the Event Horizon Telescope
(EHT). We investigate different sources of the polarization due to axion interaction with photons and the magnetic field of SMBH. These can modify the linear and circular polarization parameters of the emitted light.  We have shown that a significant circular polarization can be produced via the photon scattering from the
background magnetic field  with axions as off-shell particles. This can further constrain the parameter space of ultralight axion-like particles and their couplings with photons. The future  precise measurements of circular
polarization can probe the features of ultralight axions   in the near vicinity of SMBH. 
}

\begin{document}
\maketitle
\flushbottom

\section{Introduction}
\label{sec:intro}
One of the most interesting candidates for dark matter (DM) is axion or axion-like particle (ALP) \cite{Preskill:1982cy,Abbott:1982af,Dine:1982ah}. The QCD axion can solve the strong CP problem as well as playing the role of DM \cite{Peccei1977,Weinberg1978,Wilczek1987,DiLuzio:2020wdo}. The impact of axion field  as a DM candidate has been  extensively considered in both astrophysical observations and ground based experiments to probe its properties \cite{Marsh:2015xka,Sikivie:2020zpn,Raffelt:1990yz,Sigl:2018fba,Raffelt:2006cw,Sigl:2017wfx}. However, till now different searches have not found any confirmed evidence of axions. The recent observation of  black hole (BH)  shadow and specially the polarization map around the supermassive black hole (SMBH) of M87* by Event Horizon Telescope (EHT) collaboration 
provides some promising  new opportunities for the physics beyond the standard model (SM)~\cite{Akiyama:2019fyp,1585327,EventHorizonTelescope:2021bee,1585318}. 
 
 Generally, the photon polarization  is altered by scattering processes from the accretion disk in the vicinity of a BH. The polarization measurements specially near to event horizon of the BH can help to establish a more clear picture of the near BH region such as the configuration of the magnetic field   \cite{EventHorizonTelescope:2021bee,1585318,1585327}. Recently, it was shown that   photon rings around Kerr BHs have  universal polarization features as a pure general relativistic effect which are independent of the frequency of photons and details of the emission model
\cite{Himwich:2020msm}. 
 Due to the intense gravitational field and instability mechanism such as  superradiance, an axion cloud may be formed around a rotating BH \cite{Dolan:2007mj,Arvanitaki:2010sy,Chen:2022nbb}. It has been shown that the EHT observations can be used in order to constrain ultralight ALPs parameter space \cite{Davoudiasl:2019nlo,Davoudiasl:2021ijv}. A cloud of ultralight bosons around BHs can be probed also through its gravitational-wave (GW) signatures \cite{Baumann:2018vus,DeLuca:2021ite} or through the study of  the dynamics  of SMBH \cite{Bar:2019pnz}. Moreover, the axion cloud surrounding a Kerr BH behaves like an optically active
medium which impacts on the polarization properties of light rays traveling through
it \cite{Chen:2019fsq,Yuan:2020xui,Chen:2021lvo,Plascencia:2017kca,Chen:2022oad,Gussmann:2021mjj}. 
 Therefore, polarization observations of the emission from the accretion disks of stellar-mass and suppermassive BHs can provide a novel way to probe 
the existence of axion field close to the event horizon of a BH. Here, we  focus on specific aspect of axion coupling to electromagnetic fields which can influence the polarization of photons around a SMBH.

A background field of axion dark matter can influence on the propagation of a linearly polarized light or  distortions
of the electomagnetic (EM) spectrum \cite{Harari:1992ea,Carroll:1989vb,Marsh:2017yvc,Gong:2016zsb,Ayad:2019hrj,Ivanov:2018byi,Mena:2013baa,Liu:2019brz,Day:2018ckv}. Several studies have been done to consider the impact of the photon-axion conversion process on the photon polarization in the presence of the astrophysical magnetic field  \cite{Das:2004ee,Das:2004qka,Wang:2015dil,Masaki:2017aea,Fortin:2018aom,Guarini:2020hps,Mirizzi:2006zy,Kelley:2017vaa,Chigusa:2019rra,Dessert:2022yqq}. The birefringence effect of axion DM on the polarization of the cosmic
microwave background (CMB) has been recently investigated   \cite{Finelli:2008jv,Fedderke:2019ajk,Mukherjee:2019dsu,Sigl:2018fba,Jain:2021shf,Fujita:2020ecn,Fujita:2020aqt,Komatsu:2022nvu,Nakatsuka:2022epj,Gasparotto:2022uqo,Nakagawa:2021nme,Jain:2021shf}. The majority of  previous works focused on the linear polarization signal. One of the important fingerprint of the axion DM field is the  time-dependency of linear polarization signal that led to proposing several promising detection schemes 
 \cite{Obata:2018vvr,Castillo:2022zfl,Liu:2019brz,Yuan:2020xui}.  The linear polarization signals can  be produced by many astrophysical processes in standard scenarios and it is essential to remove the  background polarization noises in order to find axion effects.

There are two  general mechanisms that can generate circular polarization: i) intrinsic to the emission process, or ii) due to the propagation effects through Faraday conversion of linear polarization.
There are a few propagation mechanisms to generate circular polarization from the usual standard model scattering processes in the astrophysical medium~\cite{Shakeri:2018qal,Shakeri:2017knk,Hoseinpour:2020hic,Bartolo:2019eac}. Since the SM electromagnetic interactions preserve parity symmetry, they can not simply produce the circular polarization from an initially linearly polarized light. Therefore, the  circular polarization can be served as a peculiar signature of parity-violating interaction such as  photon-axion interaction~\cite{Alexander:2019sqb,Finelli:2008jv,Alexander:2008fp}. The key point of probing the circular polarization of light due to the axion effect on light polarization through mechanism (ii) is that it can be more easily discriminated from other background noises.

This paper is organized as follows. In Sec.~\ref{sec:framework} we discuss  the axion - photon interaction and  the axion distribution around a SMBH.  
A framework to compute the Stokes parameters for polarization using the quantum Boltzmann equation is presented in Sec.~\ref{sec:evolpol}. Afterwards, we study the effect of axion and magnetic fields on the linear polarization angle. The impact of different processes imposed by axion-photon interaction to generate circularly polarized signal will be considered in Sec.~\ref{sec:cirpol}. Section \ref{sec:detpol} is devoted to the investigation of detectability of the axion polarization signal by the EHT and a new stringent constraint on the  dimensionless axion-photon coupling constant will be introduced via the  upper bound  of the circular polarization reported by the latest EHT results. Finally, we summarize our results in the last section.


\section{Axion-Photon Interaction Around a SMBH }
\label{sec:framework}
Axions and axion-like particles, in general interact with photons according to the following Lagrangian
\begin{eqnarray}\label{int}
\mathcal{L}_{a \gamma}=\frac{1}{4} g_{a \gamma   }a\left(x\right) F_{\mu  \nu }\tilde{F}^{\mu  \nu }\,,
\end{eqnarray}
where $g_{a\gamma}$ is the  effective axion-photon coupling constant, $F_{\mu \nu}=\partial_\mu A_\nu-\partial_\nu A_\mu$ is the EM tensor  expressed in terms of the photon field $A_\mu$ and $\tilde F^{\mu\nu} \equiv \varepsilon^{\mu\nu\rho\sigma}F_{\rho\sigma}$ is the dual EM field tensor and $x=(\vec{r},t)$. The axion-photon coupling $g_{a\gamma}$ and the total decay constant $f_{a}$ have the following relation
\begin{eqnarray}\label{gag}
g_{a\gamma} \equiv \frac{c_{a \gamma}}{2\pi f_a}\,,
\end{eqnarray}
where the dimensionless axion-photon coupling $c_{a\gamma}\equiv\alpha_{em}(E/N-1.92)$ with $E/N$ as a model-dependent ratio of the electromagnetic and color anomaly and $\alpha_{em}$ denotes the EM fine structure constant \cite{Kim:1986ax}. For the QCD axion one has only one free parameter since $m_{a}f_{a}\approx 5.7\times 10^{-3}\text{GeV}^{2}$ \cite{diCortona:2015ldu}.  
There are several competing processes whether in the SM or beyond it that affect polarization characteristics of photons around a SMBH. Here we mainly focus on the axion-photon interaction  and consider both propagation and scattering processes in axion electrodynamics. Depending on the appearance of the axion in the external legs of scattering amplitudes, the polarization signal can be expressed as an explicit function of the axion density profile. In this case, in order to find  the value of polarization, we need to determine  the distribution of axion field around a rotating  BH. The development of an axion cloud around the Horizon of a SMBH is either
from axion dark matter at the galactic center or from production mechanisms such as superradiance process. It has been proposed that  the axion field  can gain a large density near the event horizon of a rotating  BH  through the superradiance mechanism (see Appendix \ref{axionprofile})~\cite{Brito:2014wla}. The axion field at its maximum value can reach to $a\approx f_{a}$ known as  the saturation limit  which  for ultra-light axions with $m_{a}=10^{-20}\text{eV}$ and assuming $f_{a}=10^{16}\text{GeV}$  gives 
\begin{eqnarray}
\rho_{ a}|_{\rm max}=\frac{1}{2}m_{a}^2 a^{2} |_{\rm max}=\frac{1}{2}m_{a}^2 f_{a}^{2} \approx 3.83 \times 10^{11} \left(\frac{m_{a}}{10^{-20}\text{eV}} \right)^{2} \left( \frac{f_{a}}{10^{16}\text{GeV}} \right)^{2} \,\left[\frac{\text{GeV}}{\text{cm}^{3}}\right]\,.
\end{eqnarray}
It is worth mentioning that the production of such a large axion population numbers  would not be easily achieved in practice, and a bosenova explosion might happen which prevents superradiance from persisting~\cite{Yoshino:2012kn}. It was also shown that the formation of a soliton core of fuzzy DM around  SMBHs can  lead to dense axion populations at the galactic center~\cite{Davies:2019wgi}. 
Meanwhile, the axion density can be inferred  from  the DM density near the BH which might be at most $\sim10^{10}-10^{11}\mathrm{M}_{\odot}/$pc$^{3}$ $\sim 4\times (10^{11}-10^{12})$~GeV$/$cm$^{3}$ \cite{Barausse:2014tra,Bertone:2005hw,Merritt:2003qk,Freese:2008hb}. These densities are obviously well above the ambient DM  halo density that is around  $0.3 - 0.4$~GeV$/$cm$^{3}$. 
However, for scattering of photons from the  EM backgroud field intermediated by  virtual axions which plays a major role  in our study, the signal
 does not rely on the axion density  and instead, the EM field configuration is important. In this case the precise knowledge of the axion profile  is not essential and the ultimate observable value of polarization - as will be explained in Sec. \ref{birefax} - depends upon the
magnetic field around the SMBH.
 
 The BH accretion disks containing relativistic electron-positron pairs plus possibly a small amount of ions. 
 In the case of recent EHT observation of BH shadow~\cite{Akiyama:2019fyp,1585327,EventHorizonTelescope:2021bee,1585318}, the emission region $\sim 5r_{g}$~($r_{g}\equiv GM_{BH}$) is thought to be optically thin considering an electron density $n_{e}\sim 10^{4}$~cm$^{-3}$ \cite{Akiyama:2019fyp}. The propagation length of photons around the SMBH is about   the gravitational radius  $r_{g}$ which defines as
\begin{align}\label{e29}
r_{g}\equiv G M_{BH} = 
 2.97\times 10^{-4} \text{pc}  \left(\frac{M_{BH}}{6.2\times10^{9}M_{\bigodot}} \right)\,.
\end{align}
The emission  arises close to the innermost stable circular orbit (ISCO) of the SMBH accretion disks where the magnetic field strength can be estimated as a few Gauss \cite{Rees:1984si,Caiazzo:2018evl}. For M87's SMBH detected by EHT with $M_{BH}\approx 6.2\times 10^{9} M_{\bigodot}$, the value of the magnetic field at $5r_{g}$ is about $4.9\ G$, this value in addition to other parameters has been set to reproduce the observed flux
density  of the image \cite{Akiyama:2019fyp}.

 The magnetized plasma within the plane of the accretion disk around SMBH  at low frequencies (e.g. radio waves) is birefringent giving rise to a rotation of the photon polarization position angle called Faraday rotation effect. The photons may experience similar effects but from another sources in quantum electrodynamics (QED)  such as  virtual $e^{+}e^{-}$ pairs or virtual axions  as vacuum polarization effects   \cite{Meszaros:1979xf,Shakeri:2017knk,Shakeri:2017iph,Zarei:2019sva}.  The  QED contribution to the magnetic birefringence signal scales as $\alpha B^{2}/m_{e}^{4}$ 
\cite{Shakeri:2017knk,Shakeri:2017iph}
while axion contribution scales as $g_{a \gamma }^{2}B^{2}/m_{a}^{2}$ \cite{Bogorad:2019pbu,Shakeri:2020sin,Zarei:2019sva,Evans:2018qwy}, in order to have dominant axion contribution one needs
\begin{eqnarray}\label{gm1}
\left(\frac{g_{a \gamma}}{m_{a}}\right) \gtrsim \mathcal{O}(1)\times \left(\frac{\alpha}{m_{e}^{2}}\right)\simeq 10^{-4} [\text{GeV}]^{-1}[\text{eV}]^{-1} .
\end{eqnarray}
%
The  plasma induced  polarization effect scales as $\omega_{\rm p}^{2}/\omega_{\gamma}^{2}$, where the plasma frequency $\omega_{\rm p}$ is given by
\begin{eqnarray}
\omega_{\rm p}= \left( \frac{4\pi e^{2}n_{e}}{m_{e}} \right)^{1/2}\simeq 0.56 \left(\frac{n_{e}}{10^{4}cm^{-3}}\right) [\text{GHz}]\,,
\end{eqnarray}
Note that for a physical photon in the vacuum $k^2=0$ but for a photon in plasma we should consider the effective photon mass  $k^2=\omega_{\rm p}^2$. In order to have dominant axion effects over those from plasma one needs $g_{a \gamma }^{2}B^{2}/m_{a}^{2}\gtrsim \omega_{\rm p}^{2}/\omega_{\gamma}^{2}$ which leads to the following condition
\begin{eqnarray}
\left(\frac{g_{a \gamma }}{m_{a}}\right)\gtrsim 0.2\left( \frac{\text{4.9\ G}}{B}\right)\left( \frac{n_{e}}{10^{4}cm^{3}}\right)
\left(\frac{230\text{GHz}}{\nu_{\gamma}}\right)[\text{GeV}]^{-1}[\text{eV}]^{-1}\,,
\end{eqnarray}
where  $\nu_{\gamma}=\omega_{\gamma}/2\pi $ is the frequency of the observed photon $\nu_{\gamma}$ which scaled with the characteristic frequency of EHT telescope. In order to  separate the axion effect on the polarization from the plasma effect as a frequency dependent phenomena, one can perform multi-wavelength polarimetric measurements. The above discussion has shown that for a wide range of $m_{a}$ and $g_{a\gamma}$,  axion polarization induced effects are  dominant processes, and we can explore a broad range of the axion parameter space while remaining above the SM backgrounds.



%
\section{Evolution of the Polarization and Stokes Parameters}
\label{sec:evolpol}
The polarization properties of an EM radiation can be described in terms of the Stokes parameters: I (the total intensity ),  Q and U (linear polarization), and  V (the circular polarization):
\begin{eqnarray}
\label{ipol}
I
&\equiv& \langle\dot A_x^{2}\rangle + \langle\dot A_y^{2}\rangle \,,\\ 
Q
&\equiv&  \langle\dot A_x^{2}\rangle -\langle\dot A_y^{2}\rangle \,,\\ 
U
&\equiv& \langle2\dot A_x \dot A_y \cos{(\phi_x-\phi_y}) \rangle \,,\\ V
&\equiv& \langle2\dot A_x \dot A_y \sin{(\phi_x-\phi_y}) \rangle \,,
\label{vcirc}
\end{eqnarray}
where $A$'s are the vector potential components and $\phi$'s are the phase of fields.
These equations   are defined for a monochromatic EM wave propagating in the $\hat z$ direction and electric field ($\vec E=\vec{\dot A}$) oscillates in the x-y plane and presents as 
\begin{align}
\vec E = E_{x} \hat x +  E_{y} \hat y = ( E_{0x} e^{i\phi_{x}} \hat x +  E_{0y} e^{i\phi_{y}}  \hat y) e^{-i\omega t}\,.
\end{align}
An ensemble of photons can be usually described by a  density matrix which encodes the information of intensity and the polarization of the photon ensemble and its dependency on the Stokes parameters is given by \cite{Shakeri:2017iph,Shakeri:2017knk,Shakeri:2018qal} 
\begin{eqnarray}
\rho=\frac12 \begin{pmatrix}
I+Q & U-iV\\
U+iV & I-Q
\end{pmatrix}\,.
\end{eqnarray}
The time evolution of Stokes parameters in a quantum-mechanical description of polarization is given by quantum Boltzmann equation
\begin{align}\label{bol}
{\footnotesize (2\pi)^{3}2k^{0}\delta^{(3)}(0)\frac{d}{dt}\rho_{ij}(\mathbf{k})=i\langle[\mathcal{\hat H}_{int} (t),\mathcal{\hat {D}}_{ij}(\mathbf{k})]\rangle-\frac{1}{2}\int^{+\infty}_{-\infty} dt \left<\left[\mathcal{\hat H}_{int} (t),[\mathcal{\hat H}_{int} (0),\mathcal{\hat {D}}_{ij}(\mathbf{k})]\right]\right>
\,,} 
\end{align}
where $\mathcal{\hat H}_{int}$  is  the interaction Hamiltonian and $ \mathcal{\hat {D}}_{ij}(\mathbf{k})=\hat a_{i}^{\dag}(\mathbf{k})\hat a_{j}(\mathbf{k})$ is the photon number operator. The first term on the right-hand side describes the variation of the photon polarization in a forward scattering process while the higher order terms known as  collision terms related  to the scattering processes in which the momentum of photons can be changed.  Within this quantum mechanical approach, the  quantum gauge field $A_\mu$ is a linear combination of creation  and annihilation operators as
\begin{eqnarray}
A_\mu(x) \equiv A_{\mu}^{+} (x) + A_{\mu}^{-} (x) = \int \frac{d^3 {\bf k}}{(2\pi)^3
2 \omega_{\gamma}} \left[ \hat{\bf{a}}_i({\bf k}) \epsilon _{i\mu}({\bf k})e^{-ik\cdot x}+
\hat{\bf{a}}_i^\dagger ({\bf k}) \epsilon^* _{i\mu}({\bf k})e^{ik\cdot x}
\right],
\end{eqnarray}
where $\hat{\bf{a}}_i^\dagger ({\bf k})$  and  $\hat{\bf{a}}_i({\bf k})$   satisfy the canonical commutation relation as
\begin{equation}
\left[  \hat{\bf{a}}_i ({\bf k}), \hat{\bf{a}}_j^\dagger ({\bf k}')\right] = (2\pi )^3
2\omega_{\gamma}\delta_{ij}\delta^{(3)}({\bf k} - {\bf k}' ).
\label{comm}
\end{equation}
Instead of quantum Boltzmann equation, one may consider the evolution equation for the mode amplitude in order to derive the evolution equation for the Stokes parameters. The classical methods have relied on solving the transfer equation of the polarization modes or Stokes parameters when the approximations of geometrical optics are used and  fields  are considered as classical fields \cite{Wang:2009ci,Meszaros:1979xf}.

\subsection{Propagation of Photons in an Axion Background}\label{axback}

Here we intend to consider the polarization signature of an axion background on the linearly polarized light. Assuming the axion cloud is highly non-relativistic and moves usually with  the virial velocity of galaxies which is $\sim \mathcal{O}(10^{-3})$, we can safely neglect the spatial derivative of  axion background and rewrite the interaction of Eq.~(\ref{int}) as \footnote{The photon dispersion relation in a axion background can be identified as 
\begin{eqnarray}
k^{2}=\omega_{\gamma}^{2}-|\vec k|^{2}\approx \pm g_{a\gamma}\left[ \omega_{\gamma}\dot a-\vec k \cdot \vec\nabla a 
\right].
\end{eqnarray}}
\begin{eqnarray}\label{int2}
\mathcal{L}_{a \gamma}=
  g_{a \gamma   }\dot a\left(\vec{r},t\right)A_{i} \varepsilon ^{ijk}\partial_{j}A_{k}+\mathrm{(total \ derivative)}\,.
\end{eqnarray}
Therefore, assuming a  time-dependent axion background 
 regarding the above form of the axion-photon interaction one can find \cite{Shakeri:2017iph,Shakeri:2017knk}
\begin{eqnarray}
\frac{d}{dt} \rho_{i j}&=&g_{a \gamma  }\dot a\left(\vec{r},t\right)\sum _{rr'}\varepsilon ^{i j k}k_j\epsilon _{i r}\epsilon _{kr'}^*\left(\rho _{j r}\delta _{r'i}-\rho _{ir'}\delta _{j r}\right)\\ \nonumber
&=&g_{a \gamma   }\dot a\left(\vec{r},t\right)\underset{\text{r=1,2}}{\sum }\hat\epsilon _r.\left(\hat k\times \hat \epsilon _{i}\right)\rho _{jr}-\hat\epsilon _j.\left(\hat k\times \hat \epsilon _{r}\right)\rho _{ir}\,.
\end{eqnarray}
This leads to the following set of the Stokes Parameters 
\begin{eqnarray}
\dot{I }=0\,,
\end{eqnarray}
\begin{eqnarray}\label{q1}
\dot{Q }=-2g_{a \gamma   }\dot a\left(\vec{r},t\right)U\,,
\end{eqnarray}
\begin{eqnarray}\label{u1}
\dot{U}=2g_{a \gamma   }\dot a\left(\vec{r},t\right) Q\,,
\end{eqnarray}
\begin{eqnarray}
\dot{V }=0\,.
\end{eqnarray}
By neglecting the second order time derivative of $ a\left(\vec{r},t\right)$, we can combine Eqs.~(\ref{q1}) and (\ref{u1}) as 
  \begin{eqnarray}
\ddot{Q}+\Omega^{2}Q=0,\ \ \ \ \ \ \ \ddot{U}+\Omega^{2}U=0\,,
\end{eqnarray}
where $\Omega=2g_{a \gamma}\dot a$ and these  are  simple harmonic equations.  Polarization angles $\psi$ can be defined in terms of Stokes parameters $U$ and $Q$ as
\begin{eqnarray}
\psi=\frac{1}{2}\arctan{\frac{U}{Q}}\,.
\end{eqnarray}
This electric-vector position angle (EVPA) gives valuable information about the polarization of the source. Starting with a pure linearly polarized light at the $Q$ direction ($Q_{0}=I$ and $U_{0}=0$) gives
\begin{eqnarray}
Q(t)=\cos{\Omega t}, \ \ \ U(t)=\sin{\Omega t}\,.
\end{eqnarray}
Then $\psi=\Omega  t/2$ and the variation of the polarization angle can be cast into 
\begin{eqnarray}\label{psi19}
\Delta \psi=\frac{\Omega \Delta t}{2} =g_{a  \gamma }\Delta a\left(\vec{r},t\right)\,=g_{a \gamma  } \left[ a\left(\vec{r}_{o},t_{o}\right)-a(\vec{r}_{e},t_{e})\right]\,.
\end{eqnarray}
The variation of the polarization angle $\Delta \psi$ depends on the axion density profile which can be changed between the emission point $\left(\vec{r}_{e},t_{e}\right)$
 and  the observation point $\left(\vec{r}_{o},t_{o}\right)$. In fact, the clumpy structure of axions at the middle of light ray paths  do not produce any spatial birefringence where $ a\left(\vec{r}_{o},t_{o}\right)=a(\vec{r}_{e},t_{e})$
\cite{Chen:2019fsq,Blas:2019qqp,Liu:2019brz,McDonald:2019wou,Fujita:2018zaj,Ivanov:2018byi,Plascencia:2017kca}. If the axion density at the observation point is much less than its value at the emission point, we have
\begin{eqnarray}\label{e20}
\Delta \psi\approx -g_{a \gamma   } a(\vec{r}_{e},t_{e})\,.
\end{eqnarray}
Since we  consider photons emitted from the accretion disk near the horizon of a BH where the axion field density is assumed to be very large due to superradiance mechanism,   it is reasonable to neglect $a(\vec{r}_{e},t_{e})$ in favor of $a\left(\vec{r}_{o},t_{o}\right)$. Here the axion field behaves as an optically active medium which can induce a phase shift between different polarization modes
of a linearly polarized EM field. This phase shift leads to a rotation  of the polarization plane as in Eq.~(\ref{psi19}). 
If we consider a coherently oscillating axion field   
\begin{eqnarray}\label{e411}
 a(\vec r , t)\equiv
\Re e \left(a_{0} (r) \ e^{i( {\bf p}\cdot {\bf r} -\omega t)}\right) \equiv a_{0}(r)\cos{\left[m_{a}t+\delta(r)\right]}\,,
\end{eqnarray}
 where $\delta$ is a random phase and $a_{0}$ is the axion field amplitude and the  axion field's oscillation frequency is approximately given by $m_a$  ($\omega_R\simeq m_a$ see Appendix \ref{axionprofile}).  The maximum value of axion field at the saturation limit is \cite{Yoshino:2012kn} 
\begin{align}
 \left(\frac{a_{0}}{f_{a}}\right)_{\mathrm{max}}\sim 1\,.
\end{align}
The  oscillation period is introduced by the inverse axion mass as 
\begin{eqnarray}
T\simeq\frac{2\pi}{m_{a}}\simeq4.2\times 10^{-9} \left( \frac{10^{-6}\text{eV}}{m_{a}}\right)s\,.
\end{eqnarray}
In principle the rotation angle can oscillate with the same frequency as the axion field  providing a way to measure the axion mass from the polarization of photons \cite{Fujita:2018zaj,Castillo:2022zfl,Liu:2019brz,Yuan:2020xui}. As it is shown in Eq.~(\ref{e20}), despite of the Faraday effect, the axion effect depends on the value of the axion field corresponding to the local axion density. In order to compute $\Delta \psi$ around a SMBH, we need the distribution of axion cloud around the BH which is given by \cite{Plascencia:2017kca} (see Appendix \ref{axionprofile} for more details)
\begin{eqnarray}
\label{e14}
a(t,r,\theta,\phi)=\sqrt{\frac{3M_{\rm S}}{4\pi\times 24 M_{\rm BH}}} (M_{\rm BH}m_a)^2 \mathtt{g}(\tilde{r})^{n=0, \,l=1 } \cos(\phi-\omega_R t)\sin\theta\,,
\end{eqnarray}
where $M_{\rm S}$ is the total mass of scalar particles surrounded the black hole, and $\mathtt{g}(\tilde{r})$ is the radial wave-function. The maximum value of this axion field is given by
\begin{eqnarray}\label{e3.26}
a(t,r,\theta,\phi)\Big|_{\rm max}\approx5.15\times 10^{-2} f_{a}  \cos(\phi-m_{a} t)\sin\theta\,,
\end{eqnarray}
here we assume that $\omega_R\approx m_{a}$, this leads to 
\begin{eqnarray}
\Delta \psi|_{\rm max}\approx -8.20\times 10^{-3} c_{a\gamma}  \cos(\phi-m_{a} t)\sin\theta\,, 
\end{eqnarray}
where we use Eq.~(\ref{gag}) and we consider polarized photons emitted at $(t,r,\theta,\phi)$ in Boyer-Lindquist coordinates of the BH. The maximum value of the position angle at $\theta=\pi/2$ in the plane of
the SMBH accretion disk variates between $\pm 0.47c_{a \gamma}{}^\circ $, however,  if one assumes $a_0\approx f_a$  \cite{Yoshino:2012kn} this value reaches to $\pm 8 c_{a \gamma}{}^\circ  $ \cite{Chen:2019fsq}.
We see that the rotation of polarization angle due to the axion cloud surrounding a BH is comparable to the Faraday rotation induced by the magnetized plasma pairs around the BH \cite{Sharma:2007nx,Kuo:2014pqa}.

\subsection{Axion-photon Conversion in
a Magnetic Field}
\begin{figure}[!ht]
  \centering
 \subfloat{
    \centering
    \includegraphics[width=2in]{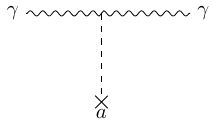} 
    \hspace{.5in}
    \includegraphics[width=2in]{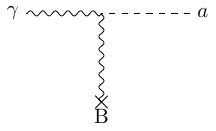}
    } 
    \caption{ Feynman diagrams showing axion-photon interaction  at tree level, (left) photon propagation in a background axion field (right) photon-axion conversion in a background magnetic field. }
     \label{background}
\end{figure}

In the presence of a background magnetic field, the photon polarization mode parallel to the magnetic field can be partially converted to axions while the amplitude of the perpendicular mode is unaffected.
In fact the external magnetic field introduces a preferred direction leading to a dichroism property  for the medium \cite{Zavattini:2012zs}. 
Due to the axion-photon conversion process, the number of photons is no longer
 conserved in the direction parallel to the magnetic field  and the angle of the linear polarization plane can be changed as
\begin{eqnarray}
\Delta \psi=\frac{1}{4}P_{\gamma\rightarrow{a}}\sin 2\alpha,
\end{eqnarray}
where $\alpha$  is
the angle of the plane of polarization with the magnetic field direction and  the photon to axion conversion probability
$P_{\gamma\rightarrow{a}}$  in a general magnetic field can be computed using perturbation theory in quantum mechanics as done 
in Refs.~\cite{Raffelt:1987im,Dessert:2019sgw,Bogorad:2019pbu,Ioannisian:2017srr}. Axion-photon conversion rate  in the presence of a uniform magnetic field is given by
\begin{eqnarray}
P_{\gamma\rightarrow{a}}=(\Delta_{M} r)^{2}\left(\frac{\sin(\Delta_{osc}r/2)}{\Delta_{osc}r/2} \right)^{2}\,,
\end{eqnarray}
here $r$ is the radial distance. We   have also introduced
\begin{eqnarray}
\Delta_{osc}=\sqrt{(\Delta_{a}-\Delta_{p}+\Delta_{QED})^{2}+(2\Delta_{M})^{2}}\,,
\end{eqnarray}
where
\begin{eqnarray}
\Delta_{M}=\frac{g_{a \gamma   } B}{2}\simeq 1.52  \left( \frac{g_{a \gamma   }}{10^{-12}\mathrm{GeV}^{-1}} \right) \left( \frac{B}{1G}\right) \text{pc}^{-1}\,,
\end{eqnarray}
\begin{eqnarray}
 \Delta_{a}=\frac{m_{a}^{2}}{2\omega_{\gamma}}\simeq 2.23\times 10^{-27} \left( \frac{m_{a}}{ 10^{-20}\text{eV}}\right)^{2} \left( \frac{\text{230GHz} }{\nu_{\gamma}}\right)\text{pc}^{-1}\,,
\end{eqnarray}
\begin{eqnarray}
\Delta_{p}=\frac{\omega_{p}^{2}}{2\omega_{\gamma}}\simeq 3.06\times10^{-9}
\left( \frac{\text{230GHz} }{\nu_{\gamma}}\right) \left( \frac{n_{e}}{10^{4}\text{cm}^{-3}}\right)\text{pc}^{-1}\,,
\end{eqnarray}
\begin{eqnarray}
\Delta_{QED}=\frac{7\alpha \omega_{\gamma}}{90\pi }  \left( \frac{B}{B_{c}} \right)^{2}\simeq 1.38 \times 10^{-11} \left( \frac{\nu_{\gamma}}{\text{230GHz}}\right) \left( \frac{B}{\text{G}} \right)^{2} \text{pc}^{-1}\,,
\end{eqnarray}
where $B_c=m_e^2/e=4.42\times 10^{13} G$ is the  QED critical magnetic field.
For these sets of scaled parameters that fit to the region around the SMBH in $\text{M87}^{\ast}$    ($\Delta_{osc}\approx 2\Delta_{M}$), in this condition the photon-axion conversion is the most effective process and the conversion probability has the maximum value as
\begin{eqnarray}\label{e22}
P_{\gamma\rightarrow{a}}\approx\sin(\Delta_{M} r)^{2}\,. 
\end{eqnarray}
Regarding this and for SMBH in $\text{M87}^{\ast}$, the value of EVPA  turns out to be
\begin{eqnarray}
\Delta \psi\approx\frac{1}{4}\sin \left(  4.51 \times 10^{-4} \left( \frac{g_{a \gamma   }}{10^{-12}\mathrm{GeV}^{-1}} \right) \left( \frac{B}{1G}\right) \frac{r}{r_g}  \right)^{2}\sin 2\alpha,
\end{eqnarray}
here $r_{g} = 2.97\times 10^{-4} ~\text{pc} $ and   $r\sim 5r_{g}$ where the dominant emission comes from, assuming $B=1 G$ and $\alpha=\pi/4$  the maximum value of polarization angle
\begin{eqnarray}\label{e28}
\Delta \psi|_{\rm max}\approx 7.28\times 10^{-5}\, ^\circ\ \left(g_{a \gamma   }/10^{-12}\mathrm{GeV}^{-1}\right)^{2}\,.
\end{eqnarray}
Assuming $f_{a}=10^{15} GeV$ and using Eq.~(\ref{gag}), the above expression for the maximum polarization angle can be re-written as  
\begin{eqnarray}\label{e225}
\Delta \psi|_{\rm max}\approx 1.16\times 10^{-8}c_{a\gamma}^{2}{}^\circ \,.
\end{eqnarray}
In addition to the variation of the EM amplitude in the direction of the magnetic field, the photon-axion mixing can introduce a phase shift between different components of the EM field \cite{Raffelt:1987im,Maiani:1986md,Sikivie:2020zpn}. Assuming a magnetic field in the $\hat{x}$ direction, $A_{x}$ component of the EM field will be transformed as $A_{x}\rightarrow{(1-P_{\gamma\rightarrow{a}}/2+i\phi)A_{x}}$, while  orthogonal component to the magnetic field $A_{y}$ remains unchanged. This induced phase, in principle can generate elliptically  polarized light due to the axion-photon interaction (see right panel of  Fig.~\ref{background}) as $\xi = |\phi|\sin2\alpha/2$ \cite{Sikivie:2020zpn,Payez:2011sh,2010A&A}. However, in our case with the maximum axion-photon conversion probability (\ref{e22}),  the light acquires negligible amount of ellipticity $\xi$ due to the axion-photon mixing (one may also check $\Delta_{osc}\approx 2\Delta_{M}$ in Eq.~(26) of Ref.~\cite{Masaki:2017aea}). Note that for $|A_{y}|=|A_{x}|$ the elliptically polaized light becomes circularly polarized. We will see in Sec.~\ref{birefax} that  the photon scattering from a magnetic field can generate sizable amount of the circular polarization in the presence of {\it{off-shell}} axions.

\section{The Generation of Circular Polarization due to Axion-photon Interaction}
\label{sec:cirpol}
The axion-photon interaction  can produce circular polarization in three  ways which will be discussed in the following sections. We will also explain the chance of detection of axion signatures by measuring the circular polarization.

\subsection{Axion Induced Propagation Effect}
\label{axphofirst}
 The polarization of EM wave can be changed by propagation in the presence of an axion background, since the photon dispersion relation is modified by the axions (See Feynman diagram of Fig.~\ref{background}). Different polarization modes of photon experience different phase velocities during propagation through an  axion cloud.
The modified dispersion relation of photon  polarization modes  is given by \cite{Alexander:2019sqb,Finelli:2008jv}
\begin{eqnarray}\label{dis}
\ddot{A}_{\pm }+\left[k^{2}\pm g_{a \gamma} \dot{a}\left(\vec{r},t\right)k \right]A_{\pm }=0,
\end{eqnarray}
where ${A}_{+} ({A}_{-})$ corresponds to the  right(left)-handed component of vector potential of a photon. Applying adiabatic condition for the variation of the axion field, the solution of Eq.~(\ref{dis}) is as follows
\begin{eqnarray}\label{aplus}
{A}_{\pm}={A}^{0}_{\pm}e^{ik\int \omega_{\pm}(t) dt}\,.
\end{eqnarray}
In the case of slowly varying $\dot{a}\left(\vec{r},t\right)$
\begin{eqnarray}\label{e44}
\omega_{\pm}(t) =\sqrt{1\pm\frac{g_{a  \gamma }\dot{a}}{k}}\,,
\end{eqnarray}
hereafter we assume the real value for $\omega_{\pm} \in \mathbb{R}$. The amplitude of the electric field components $E_{\pm}$ are proportional to $\dot{A}_{\pm}$, according to Eq.~(\ref{vcirc}) in terms of right (left) handed basis is 
\begin{eqnarray}
\label{e54}
V\equiv \langle\dot A_+^{2}\rangle - \langle\dot A_{-}^{2} \rangle \,,\\ \ \ \ \ \ \ \ \, 
I\equiv \langle\dot A_{+}^{2}\rangle + \langle\dot A_{-}^{2} \rangle \,,
\end{eqnarray}
where $\dot{A}_{\pm}$ is given by
\begin{eqnarray}\label{e55}
| \dot{A}_{\pm}| = | k\omega_{\pm} {A}^{0}_{\pm}|\,
\end{eqnarray}
in the absence of the axion-photon mixing $\omega_{\pm}=1$  and  therefore $| \dot{A}^{w/o}_{\pm}| = | k{A}^{0}_{\pm} | $ where we  defined ${A}^{w/o}$ as the photon vector potential in the case without ($w/o$) axion-photon interaction.  Regarding Eq.~(\ref{e55}), it turns out that $| \dot{A}_{\pm}| = |\omega_{\pm}  \dot{A}^{w/o}_{\pm}|$. Substituting this result into Eq.~(\ref{e54}), we find that
\begin{eqnarray}
V &=&  \omega_{+}^{2} |\dot{A}^{w/o}_{+}|^2 - \omega_{-}^{2} |\dot{A}^{w/o}_{-}|^2 \\ \nonumber &=&\frac{1}{2} (\omega_{+}^{2}+\omega_{-}^{2}) (|\dot{A}^{w/o}_{+}|^2 -|\dot{A}^{w/o}_{-}|^2 )+ \frac{1}{2} (\omega_{+}^{2}-\omega_{-}^{2}) (|\dot{A}^{w/o}_{+}|^2 +|\dot{A}^{w/o}_{-}|^2 )\,. 
\end{eqnarray}
This means 
\begin{eqnarray}
V  &=&\frac{1}{2} (\omega_{+}^{2}+\omega_{-}^{2}) V^{w/o} +  \frac{1}{2} (\omega_{+}^{2}-\omega_{-}^{2}) I^{w/o}\,,
\end{eqnarray}
where the total intensity $I=I^{w/o}+I^{\text{axion}}$, and we neglect the axion contribution to the total intensity $I\approx I^{w/o}$. Also, we assume not to have an  initial circular polarization $V^{w/o}=0$, then 
the degree of circular polarization is given by \cite{Finelli:2008jv,Alexander:2019sqb,Shakeri:2017iph} 
\begin{eqnarray}
\label{cp}
\Pi_{\rm V}\equiv \frac{V}{I}\equiv\frac{|\dot{A}_{+}|^{2}-|\dot{A}_{-}|^{2}}{|\dot{A}_{+}|^{2}+|\dot{A}_{-}|^{2}}=\frac{1}{2} (\omega_{+}^{2}-\omega_{-}^{2})\simeq\frac{2\pi g_{a  \gamma}\dot{a}}{\omega_{\gamma}}+\mathcal{O}(g_{a  \gamma}^{2})\,.
\end{eqnarray}
As it is seen, the fraction of circular polarization   depends on the frequency of photons, axion-photon coupling and  also to the temporal and spatial variation of the axion field. By using Eq.~(\ref{cp})  and scaling the photon frequency to a specific value $\mathrm{230~GHz}$ and axion mass to $10^{-20}$ eV we obtain
\begin{align}
\Pi_{\rm V}=\frac{ 2 \pi g_{a \gamma }\dot{a}}{\omega_{\gamma}}=5.41\times 10^{-19}\ c_{a\gamma} \left(\frac{m_{a}}{10^{-20}\mathrm{eV}}\right)\left(\frac{\mathrm{230GHz}}{\nu_{\gamma}}\right)\sin(\phi-m_a t)\sin\theta\,,
\end{align}
here we used Eqs.~(\ref{e3.26}) and (\ref{gag}). If one assumes  $a_0\approx f_a$ the maximum value of circular polarization  for the above scaled values of mass and energy can be reached to $1.05\times 10^{-17}\ c_{a\gamma}$.

\subsection{Scattering of Photon from Axion}
\label{axphosecond}
\begin{figure}[!ht]
  \centering
    \subfloat{
    \centering
    \includegraphics[width=2in]{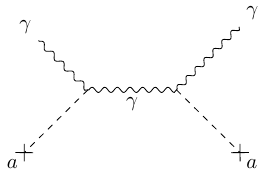} 
    \hspace{.5in}
    \includegraphics[width=2in]{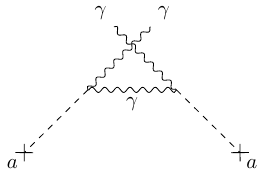}
   
    }
    \caption{Feynman diagrams showing axion photon scattering with virtual photons in the intermediate states.}
     \label{fig22}
\end{figure}

The circular polarization can also be
generated by the conversion of linear polarization through Faraday conversion due to the axion-photon scattering process. The Feynman diagrams for axion-photon scattering process at the second order of coupling   $O(g_{a \gamma }^{2}$) is shown in Fig.~\ref{fig22}.  In the following we use the quantum Boltzmann approach  to compute the evolution of the Stokes parameters  at the second order of  $g_{a \gamma }$. Let us  consider the axion field as a homogeneous plane wave as defined in Eq.~(\ref{e411})  \cite{Sikivie:2020zpn}  
\begin{eqnarray}
a(\vec r , t)=
\Re e \left(a_{0} \ e^{i( {\bf p}\cdot {\bf r} -\omega t)}\right)\,,
\end{eqnarray}
where $\omega=\sqrt{m_{a}^{2}+|{\bf p}|^{2}}$ and  $a_{0}$ is a constant factor representing the axion field amplitude. Considering this form of axion field makes the computation of this section more  feasible. Taking Eq.~(\ref{int2}) for the axion-photon interaction and
to compute the polarization effects raised from the second order effect we should use the following Hamiltonian \cite{Alexander:2008fp} %
\begin{eqnarray}
H^{(2)}_{a\gamma}&=&\nonumber-\frac{i(2g_{a\gamma})^2}{2}\int d^3 {\bf{x}} d^4 y \int \frac{d^{4}k}{(2\pi)^{4}}
\frac{d^{4}p_1}{(2\pi)^{4}}
\frac{d^{4}p_2}{(2\pi)^{4}} \varepsilon^{\mu\nu\alpha\beta} \varepsilon^{\rho\sigma\lambda\kappa} \partial_{\nu}a(x) \partial_{\sigma}a(y) (ig_{\mu\rho})\\  &\times & \Bigg[\left(\frac{e^{-ik\cdot(x-y)}}{k^2+i\epsilon}\right)\left(p_{1\alpha}(p_2-k)_{\lambda}-k_{\alpha}(p_2-k)_{\lambda}A^{-}_{\kappa}(y,p_2)A^{+}_{\beta}(x,p_1)\right)+
\\ \nonumber  && ~ \left(\frac{e^{-ik\cdot(y-x)}}{k^2+i\epsilon}\right)\left(p_{1\alpha}(p_2-k)_{\lambda}-k_{\alpha}(p_2-k)_{\lambda}A^{-}_{\kappa}(x,p_1)A^{+}_{\beta}(y,p_2)\right)
\Bigg]\,.
\label{eq:hamilsec}
\end{eqnarray}

The time evolution of the Stokes parameters  induced by axion-photon scattering can be obtained through the quantum Boltzmann equation as 
\begin{eqnarray}
\label{e12}
\dot I(\mathbf{k})=0,
\end{eqnarray}
\begin{align}
\dot Q(\mathbf{k})=\frac{4g^2_{a  \gamma}a_{0}^{2} ( p\cdot  k)}{\omega_{\gamma}}
\Big[ \frac{1}{m_{a}^{2}+2k\cdot p}-\frac{1}{m_{a}^{2}-2k\cdot p} \Big]
\Big[{\bf p}_{x}{\bf p}_{y}V(\mathbf{k})\Big]\,,
\end{align}
\begin{align}
\dot U(\mathbf{k})=-\frac{2g^2_{a  \gamma}a_{0}^{2} ( p\cdot  k)}{\omega_{\gamma}}
\Big[ \frac{1}{m_{a}^{2}+2k\cdot p}-\frac{1}{m_{a}^{2}-2k\cdot p} \Big]
\Big[({\bf p}_{x}^{2}-{\bf p}_{y}^{2})V(\mathbf{k})\Big]\,,
\end{align}
\begin{align}\label{e15}
\dot V(\mathbf{k})  =\frac{2g^2_{a  \gamma}a_{0}^{2} ( p\cdot  k)}{\omega_{\gamma}}
\Big[ \frac{1}{m_{a}^{2}+2k\cdot p}-\frac{1}{m_{a}^{2}-2k\cdot p} \Big]
\left[\left({\bf p}_{x}^{2}-{\bf p}_{y}^{2}\right) 
U(\mathbf{k})-2{\bf p}_{x}{\bf p}_{y}Q(\mathbf{k})\right]\,,
\end{align}
 where $p=(\omega_{a},{\bf p})$ and $k=(\omega_{\gamma},{\bf k})$ are axion four momentum vector and photon four momentum vector, respectively. Moreover, the wave vector $\hat{k}=(0,0,1)$ and polarization vectors $\hat{\epsilon}_1=(1,0,0)$ and $\hat{\epsilon}_2=(0,1,0)$ are assumed. Note that the polarization variation induced by axion-photon  scattering shown in the Feynman diagrams of Fig.~\ref{fig22} is a result of the spatial variations of axion field. The set of Eqs.~(\ref{e12}) - (\ref{e15}) leads  to a simple harmonic equation for the circular polarization parameter as
 \begin{align}\label{os}
\ddot{V}+\Omega^{2}V=0,
\end{align}
where the frequency of oscillation $\Omega$ defines
 \begin{align}\label{e17}
\Omega=\frac{2g^2_{a  \gamma}a_{0}^{2} ( p\cdot  k)}{\omega_{\gamma}}\left({\bf p}_{x}^{2}+{\bf p}_{y}^{2}\right) 
\Big[ \frac{1}{m_{a}^{2}+2k\cdot p}-\frac{1}{m_{a}^{2}-2k\cdot p} \Big],
\end{align}
for a non-relativistic cold axion cloud  $m_{a}\gg{\bf p}\approx 10^{-3}m_{a} $ 
\begin{align}
p\cdot  k= \omega_{\gamma}\sqrt{m_{a}^{2}+\bf p^{2}}-{\bf p\cdot \bf k} \simeq   
\omega_{\gamma} {m_{a}}\,.
\end{align}
We assume that the axion cloud  move with the velocity $ v\approx 10^{-3}$ with respect to the radiation source  \cite{Sikivie:2020zpn,Arza:2019nta}. Then the absolute value of $\Omega$ from Eq.~(\ref{e17}) becomes 
 \begin{align}
\Omega=-8g^2_{a  \gamma}a_{0}^{2}\frac{ \omega_{\gamma}}{m_{a}^{2}-4\omega_{\gamma}^{2}}\left({\bf p}_{x}^{2}+{\bf p}_{y}^{2}\right)\, ,
\end{align}
assuming ${\bf p}_{x}\simeq {\bf p}_{y}\simeq m_{a}v$ and $a_0\approx f_a$ \cite{Yoshino:2012kn} leads to
 \begin{align}
\Omega\approx1.62\times 10^{-30} c_{a\gamma}^{2}  \left(\frac{m_{a}}{10^{-20}\text{eV}} \right)^{2}\left(\frac{\mathrm{230GHz}}{\nu_{\gamma}}\right)s^{-1}\, ,
\end{align}
where Eq.~(\ref{gag}) is  used. 
Supposing a totally linearly polarized EM field originating from BH accretion disk  with  initial polarization values as $V_{0}=0$, $U_{0}=Q_{0}=\sqrt{I/2}$ the solution of Eq.~(\ref{os}) is obtained as 
\begin{eqnarray}
 \Pi_{\rm V}\equiv \frac{|V|}{I}=|\sin (\Omega  t)|\approx |\Omega | t = 4.94\times 10^{-25} c_{a\gamma}^{2} \left(\frac{r}{r_{g}}\right)\,, 
 \end{eqnarray}
 where $r_{g}$ is given by Eq.~(\ref{e29}) and we used the mass of SMBH of $\text{M87}^{\ast}$   observed by  EHT which is $M\approx 6.2\times 10^{9} M_{\bigodot}$ and for $m_a=10^{-20}~\text{eV}$ and $\nu_{\gamma}=\mathrm{230~GHz}$.

\subsection{Photon Scattering from a Magnetic Field}
\label{birefax}
\begin{figure}[!ht]
  \centering
    \subfloat{
    \centering
    \includegraphics[width=2in]{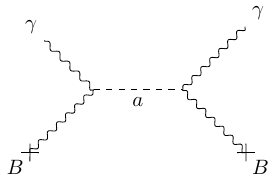} 
    \hspace{.5in}
    \includegraphics[width=2in]{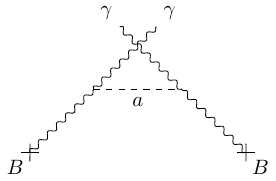}
    }
    \caption{ Feynman diagrams showing photon scattering from the background  magnetic field where virtual axion particles are in the  intermediate states.}
      \label{fig32}
\end{figure}

In this section, we study the scattering process of photons from a magnetic field by exchanging virtual axions in the intermediate states as a second order phenomena in terms of $O(g_{a \gamma }^{2}$) using the quantum Boltzmann approach. The Feynman diagram for this process is shown in Fig.~\ref{fig32}. 
The interaction Lagrangian for the  axion-photon interaction in the presence of a background magnetic field is given by
\footnote{It was recently proposed an experimental design in Ref.~\cite{Zarei:2019sva}, where in the presence of a non-uniform background magnetic field of a small cavity,  a resonant scattering process can be happened. The resonance enhances the birefringence effect which is  due to the momentum exchange between the photons and the magnetic field and consequently the  signal can be amplified with the potential to reach the sensitivity to probe axion-photon coupling at the QCD axion region. } \cite{Peccei1977,Weinberg1978,Wilczek1987} 
\begin{align}
\mathcal{L}_{\gamma a}=\frac{1}{2}g_{a \gamma }a\left(x\right) \,\varepsilon^{\mu\nu\alpha\beta} F^{B}_{\mu\nu}\partial_{\alpha}A_{\beta}~.
\end{align}
The Hamiltonian of the interaction can be defined as
\begin{align}
H^{(2)}_{\gamma B}=\frac{g^2_{a  \gamma}}{2}\int d^4x' d^3x  ~ \varepsilon^{\mu\nu\alpha\beta}\varepsilon^{\mu'\nu'\alpha'\beta'} F^{B}_{\alpha\beta} (x)F^{B}_{\alpha'\beta'} (x')D(x'-x)
\nonumber \\ \times
\left [\partial_{\mu}A^{-}_{\nu}(x') \partial_{\mu'}A^{+}_{\nu'}(x) + \partial_{\mu'}A^{-}_{\nu'}(x) \partial_{\mu}A^{+}_{\nu}(x')\right ]\,.
\end{align}
We only focus on the magnetic field components of the field strength tensor $F^{B}_{\mu \nu}$ where $B^k(x)=-\epsilon^{k i j}F^{B}_{ij}(x)/2$ is the  background magnetic field. The Hamiltonian can be rewritten as \cite{Zarei:2019sva,Shakeri:2020sin}
\begin{eqnarray}
\label{maghamil}
H^{(2)}_{\gamma B}(t) &=&\frac{g^2_{a  \gamma}}{2}\sum_{s,s'}\int \frac{d^{3}\mathbf{p}}{(2\pi)^{3}}\frac{d^{3}\mathbf{p}'}{(2\pi)^{3}}
\frac{d^4k}{(2\pi)^4}
\frac{1}{k^{2}-m_{a}^{2}+i\omega_{\mathbf{k}}\Gamma_{a}}\hat{\bf{a}}_{s'}^{\dag}(\mathbf{p}') \hat{\bf{a}}_s(\mathbf{p})
  \nonumber \\ &&
 \times \int dt' d^3x' d^3x
\Big[{\epsilon}^s\cdot\mathbf{B}(x)~
{\epsilon}^{s'}\cdot\mathbf{B}(x')
e^{-i(\mathbf{k}-\mathbf{p})\cdot \mathbf{x}}e^{i(\mathbf{k}-\mathbf{p}')\cdot \mathbf{x}'}e^{it(\omega_{\mathbf{ k}}-\omega_{\mathbf{ p}})}e^{-it'(\omega_{\mathbf{ k}}-\omega_{\mathbf{ p}'})}
\nonumber \\ &&
+{\epsilon}^{s'}\cdot\mathbf{B}(x)~{\epsilon}^{s}\cdot\mathbf{B}(x')
e^{-i(\mathbf{k}-\mathbf{p'})\cdot \mathbf{x}}e^{i(\mathbf{k}-\mathbf{p})\cdot \mathbf{x}'}e^{it(\omega_{\mathbf{ k}}-\omega_{\mathbf{ p'}})}e^{-it'(\omega_{\mathbf{ k}}-\omega_{\mathbf{p}})}
\Big]\,.
\end{eqnarray}
In the presence of a constant magnetic field without the momentum exchange we have 
\begin{align}
\label{hamilevol}
H^{(2)}_{\gamma B_{0}}=\frac{g^2_{a  \gamma}}{2}\sum_{s,s'}\int \frac{d^{3}p}{(2\pi)^{3}}\frac{1}{p^{2}-m_{a}^{2}+i\omega_{\mathbf{p}}\Gamma_{a}}\left({\epsilon}^s\cdot\mathbf{B}_{0} ~{\epsilon}^{s'}\cdot\mathbf{B}_{0} \right)\hat{\bf{a}}_{s'}^{\dag}(\mathbf{p}) \hat{\bf{a}}_s(\mathbf{p})\,.
\end{align}
The axion decay rate $\Gamma_a=\Gamma_0(a\rightarrow \gamma \gamma)+\Gamma_{B_{0}}(a\rightarrow \gamma)$ contains two terms   $\Gamma_{0}=g_{a\gamma}^2m_{a}^3/64\pi$  as the axion to two photons decay rate  and $\Gamma_{B_{0}}=P_{\gamma\rightarrow{a}}/r$ as the axion photon conversion rate in the presence of the  magnetic field \cite{Raffelt:1990yz,Zarei:2019sva}. The latter is obtained from Eq.~(\ref{e22}) which depends on the propagation length of photons in the axion cloud and can be estimated as 
\begin{eqnarray}
		\Gamma_{B_{0}}\approx\frac{g_{a\gamma}^{2}   \mathbf{B}_{0}^2 r}{4} = 1.12\times 10^{-34}  \, c_{a\gamma}^{2} \left( \frac{B_{0}}{\text{G}} \right)^{2} \left( \frac{r}{r_g} \right) \text{eV}.
\end{eqnarray}
Since $\Gamma_{B_{0}}\gg \Gamma_{0}$, one has  $\Gamma_a\approx\Gamma_{B_{0}}$, and 
working in a parameter region with $m_a\gg  (\omega_\mathbf{\gamma} \Gamma_{B_{0}})^{1/2}$  leads to

	\begin{eqnarray}\label{e5.27}	m_{a}\gtrsim3.27\times 10^{-19}   c_{a\gamma}\left(\frac{\nu_{\gamma}}{230\text{GHz}}\right)^{1/2} \left( \frac{B_{0}}{\text{G}} \right)\left( \frac{r}{r_g} \right)^{1/2} \text{eV}.
\end{eqnarray}
In this parameter space the forward scattering of photons from approximately constant magnetic background is a dominant process (compared to axion-photon conversion) and axions  appear as virtual particle in the scattering process (see Fig.~\ref{fig32}). Therefore, axions can not be popped up from the vacuum and $i \omega_\mathbf{k} \Gamma_{a}$ can be safely neglected from the denominator of Eq.~(\ref{hamilevol}). In order to compute the evolution of the Stokes parameters, we use the forward scattering term $\langle[\hat H_{int} (t),\mathcal{\hat {D}}_{ij}(\mathbf{k})]\rangle$ of the quantum Boltzmann equation (\ref{bol}) and  the below contraction relation
\begin{align}\label{e30}
\langle \hat a_{s^{'}}^{\dagger}(k^{'})\hat a_{s}(k)\rangle=2k^{0}(2\pi)^3\delta^{3}(\mathbf{k}-\mathbf{k}^{'})\rho_{s^{'}s}(\mathbf{k}).
\end{align} 
Then the time evolution of the density matrix is given by
\begin{eqnarray}
\frac{d\rho_{ij}}{dt}&=&-\frac{i\omega_{\gamma}g^2_{a  \gamma}}{m_{a}^{2}}
\sum_{\rm s} \mathbf{B}_{0}\cdot 
\epsilon^{s}
\left(\rho_{sj}\epsilon_{i}\cdot\mathbf{B}_{0}-\rho_{is}\epsilon_{j}\cdot\mathbf{B}_{0}\right)\,,
\end{eqnarray}
and the Stokes parameters can be obtained as follows
 \begin{eqnarray}
 \label{II2}
\dot I(\mathbf{k})=0 \,,
\label{I}
\label{i12}
\end{eqnarray}
\begin{eqnarray}\label{qdo}
\dot Q(\mathbf{k})=\frac{2\omega_{\gamma}g^2_{a  \gamma}}{m_{a}^{2}}B_{x}B_{y} V(\mathbf{k})
\label{Q}\,,
\end{eqnarray}
\begin{eqnarray}
\dot  U(\mathbf{k}) =-\frac{\omega_{\gamma}g^2_{a  \gamma}}{m_{a}^{2}}\left(B_{x}^{2}-B_{y}^{2}\right) V(\mathbf{k})\,,
\label{u}
\end{eqnarray}
\begin{eqnarray}\label{VV2}
\dot V(\mathbf{k}) =\frac{\omega_{\gamma}g^2_{a  \gamma}}{m_{a}^{2}}
\left[\left(B_{x}^{2}-B_{y}^{2}\right) 
U(\mathbf{k})-2B_{x}B_{y}Q(\mathbf{k})\right]\,.
\end{eqnarray}

\begin{figure}
    \centering
    {
        \includegraphics[width=5in]{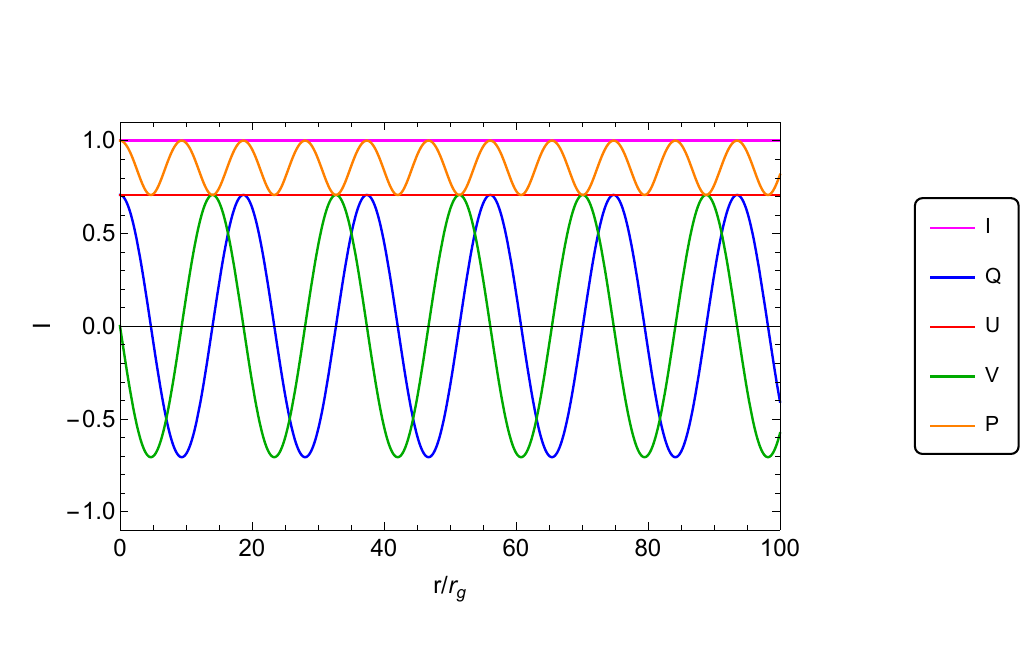}
    }
    \caption{The evolution of Stokes parameters at different distance from the center of black hole are shown in this plot. The initial conditions are $I_{0}=1$, $Q_{0}=1/\sqrt{2}$, $U_{0}=1/\sqrt{2}$ and  $V_{0}=0$. The frequency of  oscillations depends on the value of $\Omega$ which is taken from Eq.~(\ref{eq:omfig}).}\label{fig:Stokes1}
\end{figure}

As it can be seen from Eq.~(\ref{II2}), the intensity of photons
does not change during the forward scattering process from
a constant magnetic field via virtual axion exchange. The set of Eqs.~(\ref{i12}) - (\ref{VV2}) leads to a simple harmonic equation for the circular polarization parameter as
 \begin{align}
\ddot{V}+\Omega^{2}V=0\,,
\end{align}
where the frequency of the oscillation $\Omega$ is 
\begin{eqnarray}
\label{eq:omfig}
\Omega&=&\frac{\omega_{\gamma}g^2_{a\gamma }}{m_{a}^{2}}\left(B_{x}^{2}+B_{y}^{2}\right)\\ \nonumber
&\approx&3.36 \times 10^{-1} \left( \frac{\nu_{\gamma}}{230\text{GHz}}\right) \left( \frac{10^{-19}\text{eV}}{m_{a}} \right)^{2} \left( \frac{g_{a \gamma }}{10^{-17}\text{GeV}^{-1}}\right)^{2} \left( \frac{B_{0}}{1\text{G}}\right)^{2}r_g^{-1}\,.
\end{eqnarray}
By using Eq.~(\ref{gag}) and $f_{a}=10^{15}$~GeV  and reasonable scaling values for other parameters we obtain 
\begin{eqnarray}\label{e18}
 \Pi_{\rm V}\equiv \frac{|V|}{I}=|\sin (\Omega  t)|\approx |\Omega | t = 0.85 \left(\frac{c_{a\gamma}}{0.1}\right)^{2} \left( \frac{\nu_{\gamma}}{230\text{GHz}}\right) \left( \frac{10^{-19}\text{eV}}{m_{a}} \right)^{2} \left( \frac{B_{0}}{1\text{G}}\right)^{2}\left(\frac{r}{r_{g}}\right)\,.
 \end{eqnarray}
 We see that a significant amount of the circular polarization can be produced by the propagation of photons up to a large distance $r_{g}$ around a SMBH  in the presence of the magnetic field. The generation of the circular polarization in our case is due to conversion of linear polarization of the radiation as a result of the axion-photon interaction in the presence of a background magnetic field. In spite of the axion-photon scattering (in Sec.~\ref{axphosecond}),  the axion birefringence effect considered in this section is independent from the local density of axion DM and even the production mechanism of axions. 
 
 The evolution of Stokes parameters as a function of the propagation length ``r'' from Eqs.~(\ref{II2})-(\ref{VV2}) are shown in Fig.~\ref{fig:Stokes1}. All the Stokes
parameters are normalized by the intensity I, giving rise to dimensionless quantities. We supposed a totally
linearly polarized emission $P_{0}=\sqrt{Q_{0}^{2}+U_{0}^{2}}=I_{0}$  without any initial circular polarization $V_{0}=0$. 
 Assuming the set of parameters $\nu_{\gamma}=230$~GHz, $m_{a}=10^{-19}$~eV, $g_{a \gamma}=10^{-17}$~GeV$^{-1}$, and $B_{x}\simeq B_{y}\approx1$~G, we find the evolution of the Stokes parameters along the photon ring from the emission point near the SMBH up to  several gravitational radii $r_{g}$.  As it is shown, both linear and circular polarization parameters show an oscillatory behaviour with  frequency $\Omega$ which is defined by  Eq.~(\ref{eq:omfig}). The sign of
the circular polarization shows 
the handedness of the polarization, according to Fig.~\ref{fig:Stokes1} the handedness is changing with propagation length after several $r_{g}$ which can produce a unique polarimetric structure for  different photon rings. In fact  the polarization properties of different photon rings around a SMBH could depend on the number of rings analogous  to the effect presented in Ref.~ \cite{Himwich:2020msm}.
One can also consider the spatial and temporal variation (profile) of magnetic field around the BH  \cite{Day:2019bbh,Sigl:2017sew,Kelley:2017vaa}, however,  we do not consider it here to simplify the solution of integrals in Eq.~(\ref{maghamil}) and leave it for our future works. In fact, the magnetic  field strength decreases as photons propagate  farther away from the event horizon of the BH, so it can be interesting to include  the varying  magnetic
field profile with radial distance.

\section{Detectability of the Axion Polarization Signal}
\label{sec:detpol}
As mentioned earlier various experiments  can be dedicated to probe the circular polarization induced by ALPs. 
A renewed exploration of Very Long
Baseline Interferometry (VLBI) imaging strategies  taking into account polarimetric measurements is essential for extracting physical signature and cross check our  predictions. Full-Stokes polarimetry for several active galactic nuclei (AGN) 
sources with stable linear and circular polarization in the GHz regime has been presented in Ref.~\cite{2018A&A...609A..68M}, where it is shown that the reachable accuracy  in terms of polarization degree is of the order of $0.1-0.2~\%$ and the polarization angles
can be determined with an accuracy of almost $1^{\circ}$. The Atacama Large Millimeter/submillimeter Array (ALMA) using  VLBI method in 2017 April, measured  linear polarizations  in a range between $2~\%$ and $15~\%$ for several sources including Sgr $\text{A}^{\ast}$, $\text{M87}^{\ast}$  and a dozen of AGNs \cite{1585327}. The recent observations of Event Horizon Telescope can be the most promising one which has already been  measured in the first polarization map of SMBH at the centre of $\text{M87}^{\ast}$   \cite{Akiyama:2019fyp,1585327,EventHorizonTelescope:2021bee,1585318}.   
 A measurement of the linear polarization of about $\sim\%1$ for $\text{M87}^{\ast}$   at 
230 GHz has been reported previously in Ref.~\cite{Kuo:2014pqa}. Unresolved polarization 
measurements toward M87’s core from ALMA-only data shows the net polarization fraction of about $\sim 2.7~\%$. 
\begin{figure}
    \centering
    {
 \includegraphics[width=5in]{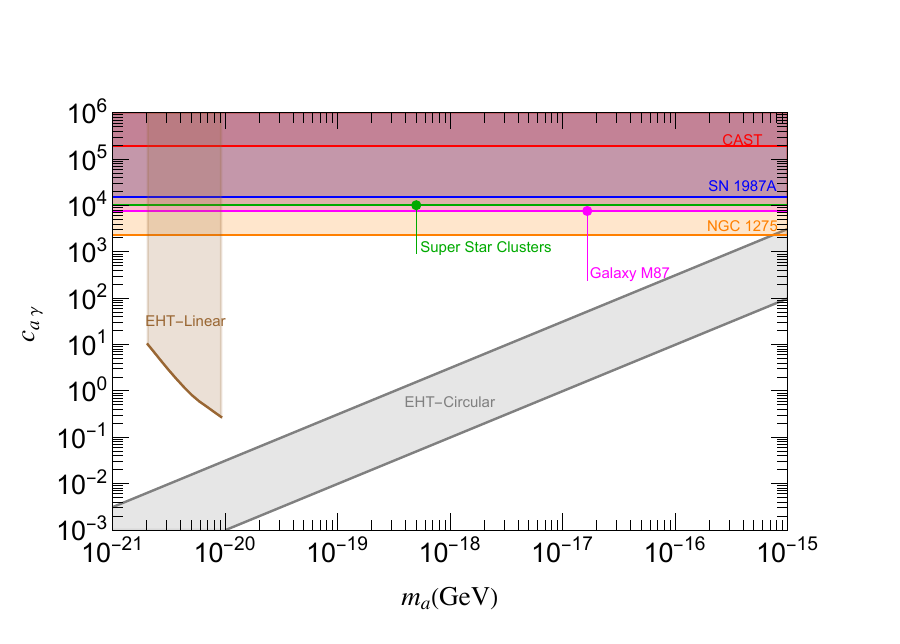}
    }
    \caption{Constraints on axion mass $m_a$ and its coupling to photons $c_{a\gamma}\equiv 2\pi g_{a\gamma}f_{a}$ where $f_{a}=10^{15}$~GeV is assumed. The bound from circular polarization of light from EHT is shown by gray color. Linear polarization of light puts a bound from EHT  presented by brown color and computed in Ref.~\cite{CAST:2017uph}. The experimental bound from CAST is shown by red color \cite{Chen:2021lvo}. Astrophysical constraints  are also shown \cite{Payez:2014xsa,Dessert:2020lil,Marsh:2017yvc,Reynolds:2019uqt,Conlon:2017ofb}.  }\label{fig:Stokes}
\end{figure}
However, the time resolved polarimetric images of this source reported by EHT gives the range of the image integrated net linear polarization fraction $1-3.7~\%$ at high angular resolution  $\sim20 \ \mu \text{as}$ in a time window of 4 days (see Table 2 of \cite{EventHorizonTelescope:2021bee}).
As it was discussed in Sec. \ref{axback}, the axion cloud around a SMBH produces a periodic oscillation pattern of the position angle for a linearly polarized light or EVPA,  with a frequency being equal to that of the axion field. Using the EHT time resolved  poalrization data of $\text{M87}^{\ast}$ which shows that a general  EVPA variation does not change significantly 
by time,  stringent constraint has been put on $c_{a\gamma}$ for the low mass ALPs
 \cite{Chen:2022oad,Yuan:2020xui}. A similar idea has been applied to constrain ALP-photon
coupling with polarization measurements of Sgr $\text{A}^{\ast}$ \cite{Johnson:2015iwg,Yuan:2020xui}. However, the detection of the time dependent linear polarization angle induced by axion field around a SMBH is challenging  in practice. This is due to the variable nature of the emission from the SMBH  accretion disk and also because of the finite spatial resolution of EHT observations
which leads to  washed out effect along the  azimuthal angle direction \cite{Chen:2019fsq,Yuan:2020xui}. These are in addition to other  astrophysical background noises whithin the emission region or outside of it which also change the EVPA.
Up to now, only few studies attempting to consider the circular polarization map around the SMBH \cite{Ricarte:2021frd,Tsunetoe:2022ktx,Tsunetoe:2020nws,Gold:2016hld}, one of the reason is that the measurement of Stokes parameter $\text{V}$  is  technically more sensitive to  calibration choices and residual errors compared to linear polarization components \cite{EventHorizonTelescope:2021bee}. From ALMA data we have only a conservative upper bound on the measurement of the circular polarization of $\text{M87}^{\ast}$ at 230 GHz which is around $0.8 \%$  \cite{Ricarte:2021frd,EventHorizonTelescope:2021bee,1585327}. Taking Eq.~(\ref{e18})  as the main contribution of axions to the circularly polarized signal and the value $8\times 10^{-3}$ as an upper observational limit, the areas that can be excluded by the present data limits is  illustrated in gray  in Fig.~\ref{fig:Stokes}. Note that the lower mass limit is defined through Eq.~(\ref{e5.27}) which indicated the validity range of our apporoximation to obtain Eq.~(\ref{e18}). We see that the exclusion region can potentially reach to smaller values of $c_{a\gamma}\sim\mathcal{O}(1)\, \alpha_{\rm EM}$ where  $\mathcal{O}(1)$ is the number of chiral fermions to induce the  dimensionless axion-photon coupling within different motivated theoretical models.
The EHT team is planing to
produce a precise circular polarization map in addition to linear polarization in future. Besides having enough sensitivity for circular polarization measurements, it is required to have sufficient temporal resolution to resolve  the oscillation signature of the polarization signal induced by the axion field.

 \begin{table}[htb]
\centering
\renewcommand\arraystretch{1.75}
\begin{tabular}
{ |>{\centering\arraybackslash}m{5cm} | >{\centering\arraybackslash}m{8cm} |}
\hline
 \multicolumn{2}{|c|}
 {\bf Polarization Angle  $\Delta \psi$ of Linearly Polarized Emission}\\[5pt]
 \hline
\bf Physical Process & \bf The Axion Polarization Angle \\[5pt]
 \hline 
  &  \\
Propagation of Photons in an Axion Background & $\Delta \psi=g_{a \gamma }\Delta a\left(\vec{r},t\right)$\\[25pt]
 \hline
  &  \\
Axion-photon Conversion in a Magnetic Field    & $\Delta \psi=\frac{1}{4}P_{\gamma\rightarrow{a}}\sin 2\alpha$\\[25pt]
 \hline
\end{tabular}

\caption{The angle of linear polarization for axion source terms from different physical processes.}\label{table1}
\end{table}
 
\begin{table}
\centering
\renewcommand\arraystretch{1.75}
\begin{tabular}{| >{\centering\arraybackslash}m{5cm} | >{\centering\arraybackslash}m{8cm} |}
 \hline
 \multicolumn{2}{|c|}{ \bf  Sources of Circular Polarization}\\[5pt]
 \hline
\bf Physical Process & \bf Form of the Axion Polarization Term\\[5pt]
 \hline
 & \\
Axion Induced Propagation Effect  & $\Pi_{\rm V}=\frac{2\pi g_{ a \gamma }\dot{a}}{\omega_{\gamma}}+\mathcal{O}(g_{a \gamma }^{2})$\\[25pt]
 \hline
  & \\
Scattering of Photon from Axion  & $\Pi_{\rm V}=8g^2_{a \gamma}a_{0}^{2}\frac{ \omega_{\gamma}}{m_{a}^{2}-4\omega_{\gamma}^{2}}\left({\bf p}_{x}^{2}+{\bf p}_{y}^{2}\right)r$\\[25pt]
 \hline
   & \\
 Photon Scattering from a Magnetic Field & $\Pi_{\rm V}=\frac{\omega_{\gamma}g^2_{a  \gamma}}{m_{a}^{2}}\left(B_{x}^{2}+B_{y}^{2}\right)r$\\[25pt]
 \hline
\end{tabular}
 \caption{Axion source terms for circular polarization from different physical processes.}\label{table2}
\end{table}

	
\section{Discussion and Conclusions}
\label{summary}
While the previous studies have mainly focused on the linearly polarized signal induced by the axion cloud (see Table.~\ref{table1}),, here we presented  an alternative novel approach of search for ALPs using the measurement of circular polarization of the light emitted from the accretion disk in the vicinity of a SMBH.
We have considered different possible processes regarding axion-photon interaction to generate the circularly polarized signal around the SMBH. 
 Our results show that depending on the density of axion dark matter around the BH, the mass and coupling of axion particles,  the magnetic field, the energy of photons and their propagation length,  there are three different cases to produce circular polarization (see Table.~\ref{table2} and Figs.~\ref{background}, \ref{fig22} and \ref{fig32}). We have seen that the dominant contribution 
 comes from the photon scattering from the background magnetic field around the SMBH with axions as {\it{off-shell}} particles. Although this circularly polarized signal does not depend on the density of axion dark matter, it can be enhanced due to the large propagation length around a SMBH that causes a unique polarimetric map of photon rings.  We showed that by using the significant  circular polarization signal introduced by scattering process in Fig.~\ref{fig32}, one 
 can put stringent
constraints on the dimensionless axion-photon coupling constant $c_{a\gamma}$.

Our study can
potentially have a significant impact  on the discovery of axions by using polarimetric measurements of EHT. Also, our predicted signal can be more easily discriminated from the background effects and is sensitive to the dimensionless axion-photon coupling constant. Although, our investigation in this paper was mainly based on some simple assumptions, it considers novel aspects of the circular polarization signals generated due to the axion photon interaction around a SBMH. Our computations might be also included into   numerical simulation codes such as {\tt IPOLE} \cite{Noble:2007zx,Moscibrodzka:2017lcu}, by modifying the corresponding radiative transfer equations taking into account our new axion source terms for the circular polarization. While the circularly polarized images of  SMBH accretion flows are less investigated than  their linearly polarized counterparts, the future more precise measurements of circular polarization thanks to the  upcoming polarimetric missions such as the next-generation EHT and space VLBI \cite{Raymond_2021,Gurvits:2022wgm} can be served as a unique opportunity to study axion-photon interaction  in the near vicinity of SMBHs.


\acknowledgments
The authors thank Roman Gold, Zhiren Wang and Avery Broderick for useful discussions and comments. 
SS thanks the organizers of "Axion Cosmology" meeting placed at MIAPP in Munich in February 2020. SS is  supported by the Munich Institute for Astro - and Particle Physics (MIAPP) which is funded by DFG under Germany's Excellence Strategy – EXC-2094 – 390783311. SS is also grateful to  Remo Ruffini for supporting his visit as Adjunct Professor at ICRANet Pescara in July 2022, where  the last part of this
work was done. FH is supported by the research grant “New Theoretical Tools for Axion Cosmology” under the Supporting TAlent in ReSearch@University of Padova (STARS@UNIPD), Istituto Nazionale di Fisica Nucleare (INFN) through the Theoretical Astroparticle Physics (TAsP) project and the Deutsche Forschungsgemeinschaft (DFG) through the CRC-TR 211 project number 315477589-TRR 211. FH is thankful to the organizers of DESY Theory Workshop 2021 in Hamburg, as well as  `` Probing New Physics with Gravitaional Waves'' Workshop 2022 at Mainz Institue for Theoretical Physics (MITP) of the Cluster of Excellence PRISMA (Project ID 39083149) in Mainz, for hospitality  and partial financial supports during the completion of this work. Also, FH thanks for the partial support by Institut Pascal at Université Paris-Saclay during the Paris-Saclay Astroparticle Symposium 2021, with the support of the P2IO Laboratory of Excellence (programme “Investissements d’avenir” ANR-11-IDEX-0003-01 Paris-Saclay and ANR-10-LABX-0038), the P2I research departments of the Paris-Saclay university, as well as IJCLab, CEA, IPhT, APPEC, the IN2P3 master projet UCMN and EuCAPT.

\appendix
\section{Axion Density Profile Around a SMBH}
\label{axionprofile}

To have a valid estimation of axion field around the black hole at the center of a galaxy one should find its evolution  around the horizon of the SMBH. Here we mainly follow Ref.~\cite{Plascencia:2017kca} for the distribution of the axion cloud around the black hole. The equation of motion  for axion field in the background with Kerr metric can be obtained as 
\begin{eqnarray}
\label{eqmotax}
\square ~ a = m_a^2 a\,.
\end{eqnarray}
This is computed by neglecting the self interaction term and assuming the potential is $m_a^2 a^2/2$.
A general solution for axion field in Eq.~(\ref{eqmotax}) is derived as \cite{Detweiler:1980uk}
\begin{eqnarray}
\label{axfield}
a(t,r,\theta,\phi)=e^{im\phi}S_{lm}(\theta)e^{-i\omega t} R_{nl}(r)\,,
\end{eqnarray}
\begin{eqnarray}
\label{radialeq}
R_{nl}(r)=A_{nl}\mathtt{g}(\tilde{r})\,,\,\,\,\,\,
\mathtt{g}(\tilde{r})=\tilde{r}^le^{-\tilde{r}/2}L_{n}^{2l+1}(\tilde{r})
\,,\,\,\,\,\,
\tilde{r}=\frac{2rM_{\rm BH}m_a^2}{l+n+1}\,.
\end{eqnarray}
In above definitions the Laguerre polynomials appear as $L_{n}^{2l+1}(\tilde{r})$. The spheroidal  harmonics are denoted by $S_{lm}(\theta)$. The azimuthal and temporal dependence are shown by $e^{im\phi}$ and $e^{-i\omega t}$, where $\omega\equiv \omega_R + i \omega_I$. In fact in the case $\omega_{R}<m_{a} \Omega_{H}$ the superradiance mechanism is triggered, where $\Omega_{H}$ is the
angular velocity of the BH event horizon \cite{Davoudiasl:2019nlo}.

\begin{figure}
    \centering
    {
        \includegraphics[width=2.6in]{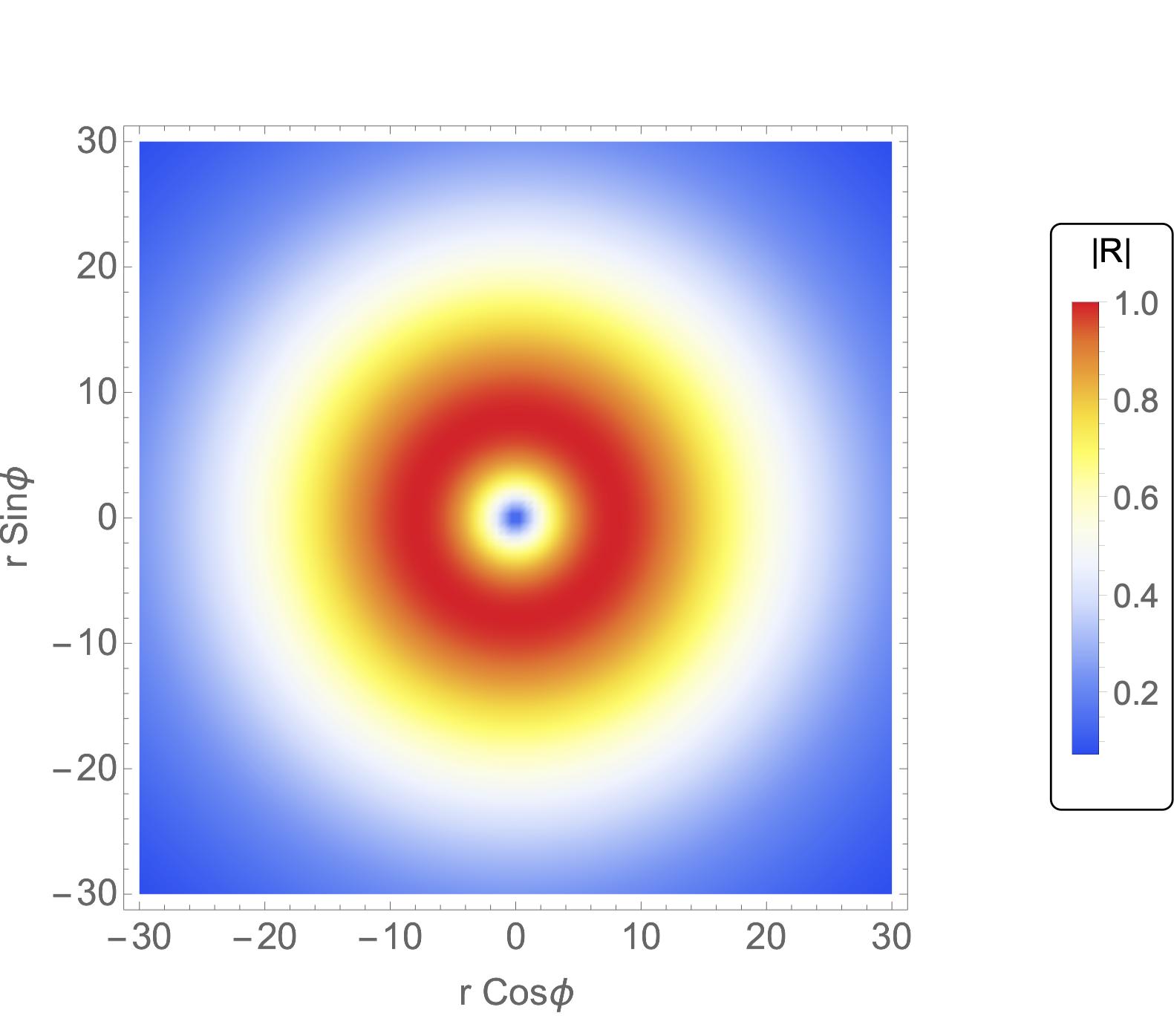}
        \label{fig:second_sub}
    }
    {
        \includegraphics[width=2.6in]{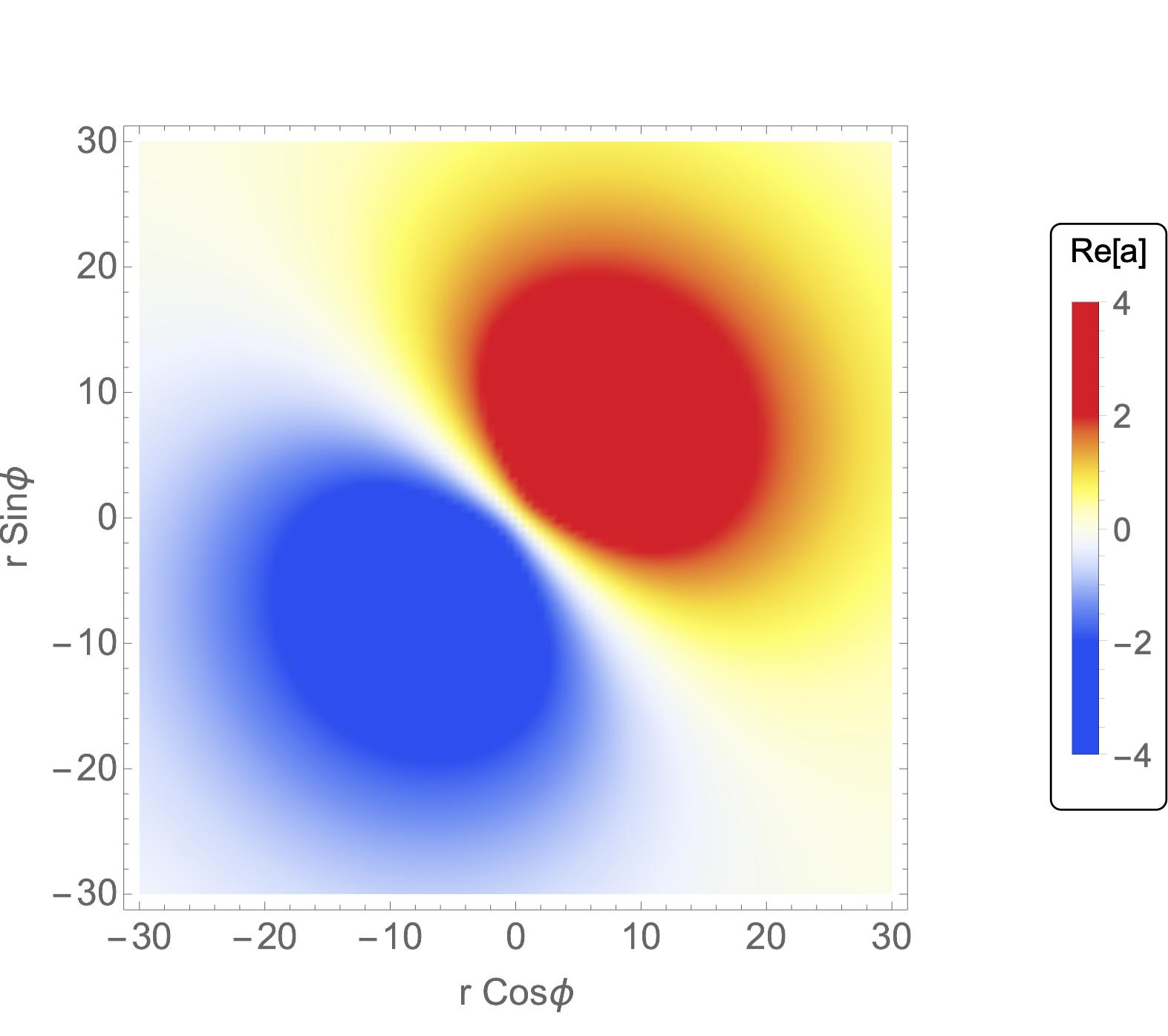}
        \label{angles}
     }
    \caption{Axion field around the  black hole is shown. The left and right panels are for the radial part of axion field  $|R|$  (scaled to its maximum) and its real part  ${\rm Re}[a]$ scaled to its numerical factor at $\theta=\pi/2$ from Eqs.~(\ref{axfield}) and (\ref{axfunc}) with $l=1$, respectively. The radial distance has been scaled to $r_{cloud}\sim8$ which means that $M_{BH}m_{a}^{2}=0.25$. }\label{fig:axbh}
\end{figure}

Here we assume the case $M_{\rm BH} m_a \ll 1~~ (l=1)$ to simplify the following quantities \cite{Plascencia:2017kca}  
\begin{eqnarray}
\omega_R= m_a-\frac{m_a}{2}\left(\frac{M_{\rm BH}m_a}{2}
\right)^2\,,
\end{eqnarray}
\begin{eqnarray}
\omega_I=\frac{\left(M_{\rm BH}m_a\right)^9}{48M_{\rm BH}}\left(\frac{J}{M_{\rm BH}^2}-2m_a r_{+}\right)\,,
\end{eqnarray}
where 
$r_{\pm}=M_{\rm BH}\left(1\pm\sqrt{1-( J/M_{\rm BH}^2)^2}\right)$ and $J$ is the spin of BH \cite{Plascencia:2017kca,Detweiler:1973zz,Detweiler:1980uk}. The radial profile of axion field for the case $l=1$ around a BH based on Eqs.~(\ref{axfield}) and (\ref{axfunc}) is shown in Fig.~\ref{fig:axbh}. The left panel is for $|R|$ and the right panel shows ${\rm Re}[a]$. 

In addition, the size of the developed axion cloud around the SMBH can be computed as \cite{Arvanitaki:2010sy}
\begin{eqnarray}
r_{cloud}\sim \frac{l+n+1}{M_{\rm BH} m_{a}^{2}}\,.
\end{eqnarray}
To find the amount of circular polarization we require to compute the time evolution of axion field using the profile defined in Eq.~(\ref{axfield}). For example we consider the case $l=m=1$ and $n=0$ as the lowest mode.

Using the temporal and spatial variation of axion field around the black hole we can compute the density of axion using the definition of energy momentum tensor. We also consider $m_a M_{\rm BH}\equiv m_a M_{\rm BH}(G/\hbar c)$. The scalar field energy density using the Kerr metric in Boyer-Lindquist \cite{Detweiler:1973zz,Detweiler:1980uk} background (coordinates) is given by \cite{Plascencia:2017kca}
\begin{eqnarray}
\label{densax1}
\rho =\frac{A_0}{2r^2}&\times&\bigg( m_a^4 M_{\rm BH}^2 r^2 \mathtt{g}^\prime({\tilde r})^2 \sin^2 \theta \cos^2[\phi-\omega_R t]+ 
\\ \nonumber 
& & ~~~
\mathtt{g}({\tilde r})^2\Big[\cos^2[\phi-\omega_R t]( \cos^2 \theta +m_{a}^2r^2 \sin^2 \theta)+\sin^2[\phi-\omega_R t] (1+\omega_R^2 r^2 \sin^2 \theta)\Big]
 \bigg)\,. 
\end{eqnarray}
Then using the density we compute the total mass of scalar particles surrounded the black hole
\begin{eqnarray}
M_{\rm S}=\int \rho ~r^2 dr ~\sin\theta ~d\theta ~d\phi\,.
\label{scmass1}
\end{eqnarray}
From Eqs.~(\ref{radialeq}), (\ref{densax1}) and (\ref{scmass1}) we obtain 
\footnote{
\begin{eqnarray}
\mathcal{I}_n=\int_0^\infty dx x^n \mathtt{g}(x)^2 \,,\,\,\,\,\,
\mathcal{I}_n^\prime=\int_0^\infty dx x^n [\mathtt{g}^\prime(x)]^2 \,,\,\,\,\,\, 
\mathcal{I}_2=\int_0^\infty dx x^2 \mathtt{g}(x)^2=24 \,.
\nonumber
\end{eqnarray}
}
\begin{eqnarray}
\label{masss}
M_{\rm S}=\frac{2\pi A_0^2}{3 M_{\rm BH} m_a^2} \left(2\mathcal{I}_0+\mathcal{I}^\prime_2+\frac{2\mathcal{I}_2}{M_{\rm BH}^2m_a^2}\right)\,,\,\,\,\,\,
A_0=\frac{3}{4\pi \mathcal{I}_2}\left(\frac{M_{\rm S}}{M_{\rm BH}}\right)\left(M_{\rm BH} m_{a}\right)^4\,.
\end{eqnarray}
Then the functionality of axion cloud can be derived \cite{Plascencia:2017kca}
\begin{eqnarray}
\label{axfunc}
a(t,r,\theta,\phi)=\sqrt{\frac{3M_{\rm S}}{4\pi \mathcal{I}_2M_{\rm BH}}}(M_{\rm BH}m_a)^2 \mathtt{g}(\tilde{r})\cos(\phi-\omega_R t)\sin\theta\,,
\end{eqnarray}
where $M_{\rm S}$ is the total mass of scalar particle (axion) cloud.

The  superradiant instability crucially depends on the dimensionless product of the BH mass and axion mass as
\begin{eqnarray}
M_{\rm BH}m_a=4.65\times10^{-1}\left(\frac{M_{\rm BH}}{6.2\times10^9 M_{\odot}}\right)\left(\frac{m_a}{10^{-20}eV}\right)\,.
\end{eqnarray}
In  the limit the axion cloud can be developed around a SMBH in a reasonable time scale due to the  super-radiance instability one obtains 
\begin{eqnarray}
\tau\approx 6.2\times10^{10}\left( \frac{M_{\rm BH}}{6.2\times10^{9}M_{\odot}}\right)\left( \frac{0.2}{M_{\rm BH} m_{a}}\right)^{9}s\,.
\end{eqnarray}

We assume that the wavelength of the EM radiation is much smaller than the  length scale over which the axion field changes. 
The maximum value of $a$ is at the order of $f_{a}$ where
the nonlinear self-interaction of the axion field becomes important, at this point the axion cloud may experience a  rapid collapse event  which is called a bosenova explosion \cite{Yoshino:2012kn}. There is an upper limit on $M_{S}/M_{\rm BH}$ corresponding to the saturation condition $a\sim f_{a}$ \cite{Plascencia:2017kca,Arvanitaki:2010sy} 
\begin{eqnarray}
\frac{M_{S}}{M_{\rm BH}}\leq \frac{2l^{4} f_{a}^{2}}{(M_{\rm BH}m_{a})^{4}}\,.
\end{eqnarray}
Here assuming $l=1$ Eq.~(\ref{axfunc}) gives
\begin{eqnarray}
\label{ax-field}
a(t,r,\theta,\phi)=\sqrt{\frac{3M_{\rm S}}{4\pi\times 24 M_{\rm BH}}} (M_{\rm BH}m_a)^2 \mathtt{g}(\tilde{r})^{n=0, \,l=1 } \cos(\phi-\omega_R t)\sin\theta\,.
\end{eqnarray}
In fact the size of the axion cloud surrounding the BH is estimated as $\tilde{r}_{\rm cloud}\sim 4$ which gives  $\mathtt{g}(\tilde{r}_{\rm cloud})\sim 0.5$.  However, the maximum value of $\mathtt{g}_{\rm max}(\tilde{r})\sim0.73$ is obtained at $\tilde{r}_{max}\sim2$, this leads to the maximum value of axion field as 
\begin{eqnarray}
a(t,r,\theta,\phi)\Big|_{\rm max}\approx5.15\times 10^{-2} f_{a}  \cos(\phi-\omega_R t)\sin\theta\,.
\end{eqnarray}

\bibliography{references}

\end{document}